\documentclass[reprint,showpacs,aps,twocolumn,pra]{revtex4-1}

\usepackage{graphicx,color}

\usepackage{amsmath}
\usepackage{amssymb}
\usepackage{amsthm}
\newcommand{\nab}{\nabla}
\newcommand{\bE}{{\bf E}}
\newcommand{\bB}{{\bf B}}
\newcommand{\bv}{{\bf v}}
\newcommand{\bk}{{\bf k}}
\newcommand{\br}{{\bf r}}

\newcommand{\bn}{\begin{enumerate}}
\newcommand{\en}{\end{enumerate}}
\newcommand{\ba}{\begin{eqnarray}}
\newcommand{\ea}{\end{eqnarray}}

\usepackage{graphicx}
\newcommand{\mj}{mJ/cm$^2$ }
\newcommand{\mje}{mJ/cm$^2$}

\newcommand{\hp}{hyperpolarizability~}

\newcommand{\an}{$\rm \AA$}

\newcommand{\be}{\begin{equation}}
\newcommand{\ee}{\end{equation}}

\newcommand{\et}{{\it et al. }}

\newcommand{\torok}{T\"or\"ok }

\def\prl{{ Phys. Rev. Lett. }}

\renewcommand{\hp}{{\bf p}}

\newcommand{\Br}{{\bf r}}
\newcommand{\rt}{({\bf r},t)}

\newcommand{\UU}{}

\begin{document}

\newcommand{\clr}{}




\title{A simple model for longitudinal electron transport during and
{\clr  after laser excitation: Emergence of electron resistive transport} }

\author{Robert Meadows}
 \affiliation{Department of Physics, Indiana State University,
   Terre Haute, IN 47809, USA}

\author{Y. Xue}
 \affiliation{Milwaukee Area Technical College, 
Milwaukee, WI 53233-1443, USA}

\author{Nicholas Allbritton}
 \affiliation{Department of Physics, Indiana State University,
   Terre Haute, IN 47809}

\author{G. P. Zhang$^*$}

 \affiliation{Department of Physics, Indiana State University,
   Terre Haute, IN 47809}

\email{guo-ping.zhang@outlook.com}

\date{\today}

\begin{abstract}
  {Laser-driven electron transport across a sample has garnered
    enormous attentions over several decades, as it provides a much
    faster way to control electron dynamics. Light is an
    electromagnetic wave, so how and why an electron can acquire a
    longitudinal velocity remains unanswered.  Here we show that it is
    the magnetic field that steers the electron to the light
    propagation direction. But, quantitatively, our free-electron
    model is still unable to reproduce the experimental
    velocities. Going beyond the free electron mode and assuming the
    system absorbs all the photon energy, the theoretical velocity
    matches the experimental observation. {\clr We introduce a concept of
    the resistive transport, where electrons deaccelerate under a
    constant resistance after laser excitation. This theory finally
    explains why the experimental distance-versus-time forms a
    down-concave curve, and unifies ballistic and superdiffusive
    transports into a single resistive transport.}  We expect that our
    finding will motivate further investigations.  }
\end{abstract}

\maketitle

\section{Introduction}

\begin{figure}
    \centering
  \includegraphics[angle=0,width=0.8\columnwidth]{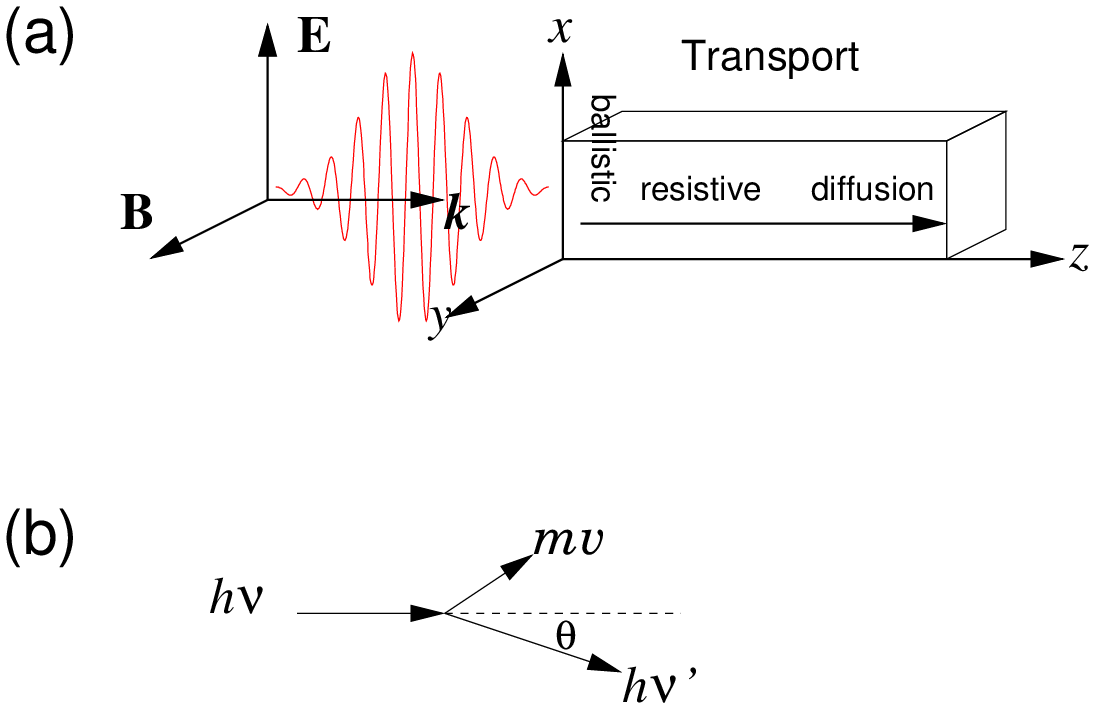}
  \caption{(a) A linearly polarized laser pulse propagates along the $z$
    axis with the wavevector ${\bf k}$, where the electric field ${\bf
      E}$ is along the $x$ axis and the magnetic field {\bf B} is
    along the $y$ axis. Electron transport is separated into two
    regimes: resistive and diffusive transports. The ballistic
    transport is in the beginning of
    the resistive transport. 
(b) Compton scattering, where a photon comes
    from the left and collides with a static electron.  }
    \label{fig1}
\end{figure}

Electron transport is a common phenomenon that occurs in various
forms. An electron can diffuse under a temperature or concentration
gradient, often described by the diffusion equation \cite{schafer},
$\frac{\partial n}{\partial t}=D \frac{\partial^2 n}{\partial
  z^2}-\frac{n}{T},$ where $n$ is the electron density, $D$ is the
diffusion constant and $T$ is the relaxation time. If an electric
field is applied, one can describe it through the Boltzmann equation
\cite{mahan2000,riffe2023}, $\frac{\partial f}{\partial t} +\bv \cdot
\frac{\partial f}{\partial \br} +\frac{e{\bf E}}{m_e}\cdot
\frac{\partial f}{\partial \bv}=-\frac{f-f_0}{T} $, where $f$ is the
electron distribution function, $\bv$ is the electron velocity, $\br$
is its position, $m_e$ is the electron mass, and ${\bf E}$ is the
electric field.  When one applies a voltage bias along the $x$ axis
across a device, the electron inside the device moves along the $x$
axis. This is how classical physics explains electron transport.  When
the device becomes shorter, one has a ballistic transport, where the
electron travels like a bullet without resistance inside the device
\cite{datta}.  Now, consider a linearly-polarized laser pulse,
propagating along the $z$ axis and with its electric field $\bE$ along
the $x$ axis and the magnetic field $\bB$ along the $y$ axis, is
applied on to the same device (Fig. \ref{fig1}). It is obvious that
the electron will move along the $x$ axis, since $\bE$ is along the
$x$. But there is no electric field along the $z$ axis. So does the
electron travel along the longitudinal direction?

The answer is affirmative.  In 1987, Brorson \et \cite{brorson1987}
employed a 96-fs laser pump pulse of wavelength 630 nm and of fluence
1 \mj to excite a group of thin gold films of thickness from 200 to
3000 $\rm\AA$ from the front, and then detected the reflected probe
beam from the back of the sample. They found that the electron can
travel at 10$^8$ cm/s, very close to the Fermi velocity of electrons
in Au.  In 2011, Melnikov \et \cite{melnikov2011} employed the same
experimental setup, but with a magnetic film Au/Fe/MgO(001), where the
thickness of Fe is 15 nm and that of Au is 50 and 100 nm.  They sent
in a $p$-polarized 35-fs pump pulse of 800 nm on to the Fe layer
through the MgO substrate, and then detected the magnetic
second-harmonic (MSH) signal reflected from the back of the Au
layer. They found the MSH hysteresis loop after 30 fs delay between
the pump and probe, even with fluence 1 \mj and energy 40 nJ per
pulse.  These results attracted a broad interest
\cite{choi2014,vodungbo2012,pfau2012,kimling2017,hofherr2017,seifert2017,seifert2018,chen2019}. Besides
MSH, Beyazit \et \cite{beyazit2020,beyazit2023} employed the
time-resolved two-photon photoemission and found that the electron
lifetimes are shorter when the thickness of their Au layer
increases. They concluded the excited electrons propagate through Au
in a superdiffusive regime \cite{battiato2012}, but ballistically
across the interfaces. The transition from a superdiffusive to a
diffusive transport occurs around 20-30 nm of the Au layer
\cite{kuhne2022}. Using the same technique, B\"uhlmann \et
\cite{buhlmann2020} detected 4\% spin polarization change in the Au
film, consistent with that of Hofherr \et \cite{hofherr2017}.
Recently, Karna \et \cite{karna2023} reported that the ballistic
length of hot electrons in gold films exceeds 150 nm, 50\% larger than
100 nm from prior studies. But their experimental geometry is
different. They measured the lateral motion (along the traverse
direction), not along the longitudinal direction
\cite{brorson1987,hohlfeld2000}.

Theoretically, several models are proposed. Salvatella \et
\cite{salvatella2016} proposed the analytical model of the
demagnetization amplitude, where they linked the absorbed energy to
the temperature change and then to the demagnetization.  Another one
is the superdiffusion model \cite{battiato2012}, where one assumes a
time-dependent diffusion exponent and then solves the diffusion
equation for the electron. These
theories build in a gradient from the beginning, but how such a
gradient is established is not addressed. This boils down to the
fundamental question:  why and how the electron could gain the
longitudinal velocity \cite{liu2023,kang2023,malinowski2018}.  In
atoms and plasma physics, a similar problem has been studied before
\cite{yang2011,maurer2021}, but with the laser fluence over $2\times
10^5$ J/cm$^2$, which has little relevance to laser-induced electron
transport in a real device.

Given the complexity of the problem \cite{jang2020}, we have two
moderate objectives.  First, we limit ourselves to a simpler question:
Is it possible to understand laser-induced electron transport
qualitatively using a free-electron model?  We employ Varga and
T\"or\"ok's method, which is based on the Hertz vector.  We carry out
a series of simulations by solving the Newtonian equation of motion of
the electron.  We find that the longitudinal velocity is a joint
effect of the electric and magnetic fields of the laser pulse.
However, using the same experimental laser parameters
\cite{razdolski2017}, we are still unable to quantitatively reproduce
the experimental results.  Going beyond the free electron model and
assuming all the photon energy absorbed into the system, we find that
the electron velocity then matches the experimental ones.  Second,
after laser excitation, we introduce a new concept of the resistive
transport, where hot electrons deaccelerate under a constant
resistance from other electrons and ions. This does not only explain
why experimentally \cite{heckschen2023,suarez1995} the dependence of
the electron time on the electron distance is a down concave, but also
unifies the ballistic transport \cite{xiong2017,suslova2018} and the
superdiffusive transport \cite{battiato2012} as a single resistive
transport.  Our finding will have an important impact on the future
investigation of ultrafast electron and spin transport
\cite{brorson1987,melnikov2011,stanciu2007,mplb18,aip20}.

The rest of the paper is arranged as follows. Section II is devoted to
our theoretical formalism. Numerical results during the laser
excitation are given in Sec. III, where we compare our results with
the Compton scattering and the experimental results. In Sec. IV, we
compare and contrast four models with the latest experimental
data. Section V introduces the resistive transport theory after the
initial excitation.  We conclude this paper in Sec. VI.

\section{Formalism}

We consider a free electron placed inside a laser field, with its
electric-field {\bf E} along the $x$ axis and the magnetic field {\bf
  B} along the $y$ axis.  Figure \ref{fig1}(a) schematically
illustrates a typical experimental geometry, where the laser pulse
propagates toward the sample along the $z$ axis.  The Hamiltonian of
the electron inside an electromagnetic field is \be
H=\frac{1}{2m}({\bf p}-q{\bf A}(\br,t))^2 +q\phi, \ee where ${\bf
  A}(\br,t)$ is the vector potential, $\phi$ is the scalar potential,
$m$ and $q$ are the electron mass and charge, respectively.  The
equation of motion is given in terms of the generalized Lorentz force
as \cite{griffiths2018},
\begin{widetext}
\be m\frac{d^2\Br}{dt^2}=q {\bf
  E}\rt+\frac{q}{2m}(\hp\times {\bf B}\rt - {\bf B}\rt \times \hp)-
\frac{q^2}{m} ({\bf A}\rt\times {\bf B}\rt). \ee
\end{widetext}
If we set ${\bf
  B}\rt$ to zero, then every term, except the first term, on the right
side, is zero, so the electron only moves along ${\bf E}\rt$. This
demonstrates that it is indispensable to include ${\bf B}\rt$.

We follow Varga and \torok \cite{varga1998}, and start from the Hertz
vector, which satisfies the vectorial Helmholtz equation, so the
resultant electric and magnetic fields automatically satisfy the
Maxwell equation, given in an integral form.  Within the paraxial
approximation, we get the reduced $\tilde{\bf E}$ and $\tilde{\bf B}$,
\be
\begin{array}{lll}
\tilde{E}_x = \left[1+ \frac{4x^2-2w_0^2}{w_0^4k^2} \right],& 
\tilde{E}_y = \frac{4xy}{w_0^4k^2 },&  \tilde{E}_z =
-\frac{2ik_zx}{w_0^2k^2 }. 
\\ \tilde{B}_x=0,&
\tilde{B}_y= -\frac{ik_z\sqrt{\epsilon\mu}}{k},& \tilde{B}_z=
-\frac{2y\sqrt{\epsilon\mu}}{kw_0^2}. 
\end{array}
\label{varga1}
\ee The final electric and magnetic fields are \be {\bf E}= A_0
\tilde{\bf E}e^{ik_zz-i\omega t-\left (\frac{x^2+y^2}{w_0^2}\right)},
      {\bf B}= A_0 \tilde{\bf B}e^{ik_zz-i\omega t -\left
        (\frac{x^2+y^2}{w_0^2}\right)}, \label{varga3} \ee where $A_0$
      is the field amplitude and $w_0$ is the pulse spatial width
      chosen as $10\lambda$. $\lambda$ is the wavelength of the pulse.
      When the light enters a sample, both fields are reduced by
      $e^{-z/(2\lambda_{pen})}$, where $\lambda_{pen}$ is the
      penetration depth and 2 comes from the fact that penetration
      depth is defined at $1/e$ the incident fluence and the fluence
      is proportional to the square of the electric
      field. $\lambda_{pen}$ is chosen to be 14 nm, typically values
      in fcc Ni.

 In our study, we choose a linearly $x$-polarized pulse that is
 propagating along the $z$ axis. The pulse is a Gaussian of duration
 $\tau$, amplitude $A_0$ and photon energy
 $h\nu$. $A_0e^{-i\omega t}$ in ${\bf E}$ and ${\bf B}$ in
 Eq. \ref{varga3} is replaced by
 $A_0e^{-t^2/\tau^2}\cos(\omega t)$, so our electric and magnetic
 fields are 
 \ba{\bf E}\rt&=&
 A_0e^{-t^2/\tau^2}\cos(\omega t) \tilde{\bf
   E}e^{ik_zz-\frac{z}{2\lambda_{pen}}-\left
   (\frac{x^2+y^2}{w_0^2}\right)},\\ {\bf
   B}\rt&=&A_0e^{-t^2/\tau^2}\cos(\omega t) \tilde{\bf
   B}e^{ik_zz-\frac{z}{2\lambda_{pen}}-\left
   (\frac{x^2+y^2}{w_0^2}\right)}.
 \label{varga4}
 \ea These analytic forms of ${\bf E}$ and ${\bf B}$, which contain
 both the spatial and temporal dependences, greatly ease our
 calculation.  As a first step toward a complete transport theory, we
 treat the electron classically, and solve the Newtonian equation of
 motion numerically, \be \frac{d{\bf v}}{dt}=\frac{q}{m}\left [{\bf
     E}(\br,t)+{\bf v}\times {\bf B}(\br,t)\right ],
\label{eq817b}
\ee where ${\bf v}$ is the electron velocity. 
Although our method is classical,
it fully embraces the real space approach that is more suitable for
transport, a key feature that is often missing from prior
first-principles and model simulations.  This will answer the most
critical question why the electron can move along the axial direction.

\section{During laser excitation}

\begin{figure}
\includegraphics[angle=0,width=0.8\columnwidth]{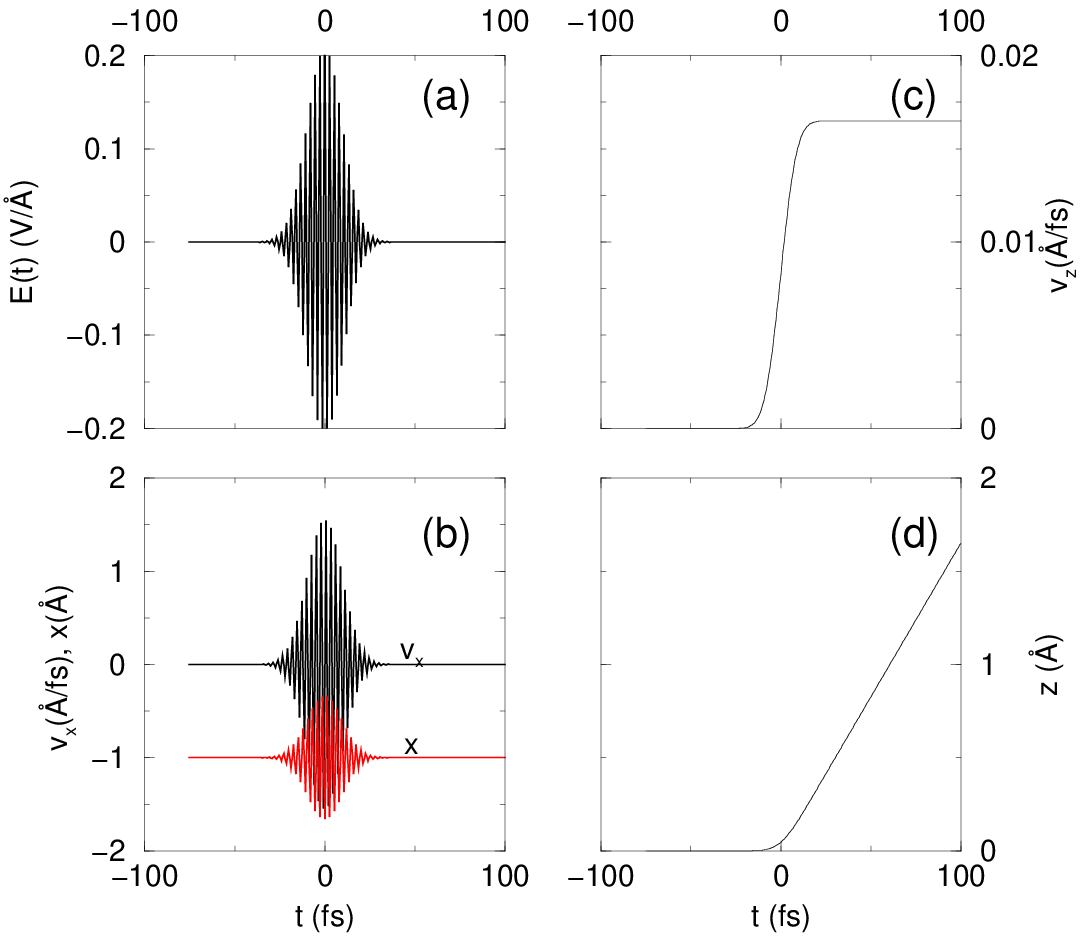}
\caption{(a) Electric field of the experimental laser pulse of 14 fs,
  fluence of 10 \mje, and photon energy of 1.55 eV
  \cite{razdolski2017}.  Here $A_0=0.207\ \rm V/\AA$.  (b) The
  electron velocity $v_x$ and position $x$ oscillate strongly with
  time.  The electron moves little along the $x$ axis. $x$ is
  downshifted by one unit for clarity.  (c) $v_z$ does not oscillate
  and only increases monotonically with time.  (d) $z$ increases with
  time $t$. }
\label{fig2}
\end{figure}

We start with the experimental laser parameters from Razdolski \et
\cite{razdolski2017}, where their laser fluence is $F=10$ \mje,
duration $\tau=14$ fs, and photon energy of $h\nu =1.55 $ eV.
These parameters are typically used in experiments \cite{jpcm10}.  We
use \cite{ourbook}
$F=\frac{1}{2}cn\epsilon_0|A_0|^2\sqrt{\frac{\pi}{2}}\tau$ to find the
electric field amplitude to be $A_0=0.207\ \rm V/\AA$.  Our electric
field is shown in Fig. \ref{fig2}(a).  The initial position and
velocity of the electron are set to zero. 
  Figure \ref{fig2}(b) shows the velocity $v_x$ and position $x$
  oscillate strongly with time. The maximum velocity reaches $\rm
  1.5\ \AA/fs$, $1.5\times 10^{5}\ \rm m/s$, 
    on the same order of
    magnitude of the velocity in \UU{ch1} 
    {\clr Co/Pt} multilayers \cite{bergeard2016}
  and also that of the experiment in Co/Cu(001) films
  \cite{wieczorek2015}. Because the kinetic energy is proportional to
  the velocity, the electron must heat up.  If we include the band
  structure of a solid, it will stimulate both intraband and interband
  transitions \cite{jpcm23}, leading to demagnetization and magnon
  generation \cite{jap19}.  We see that the final $x$ is close to
  zero, so the electron heating is local, the local heating.  For
  clarity, in Fig. \ref{fig2}(b) we shift $x$ by one unit down.

The velocity along the $z$ axis, $v_z$, is very different.  Figure
\ref{fig2}(c) shows $v_z$ is always positive, and increases with time,
without oscillation. It reaches $1.65\times 10^{-2}$ \an/fs.  The
positivity of $v_z$, regardless of the type of charge, is crucial to
the electron transport, and can be understood from the directions of
${\bf E}$ and {\bf B}.  Suppose at one instant of time, {\bf E} is
along the $+x$ axis and {\bf B} along the $+y$ axis. The electron
experiences a negative force along the $-x$ axis and gains the
velocity along the $-x$ axis, so the Lorentz force due to {\bf B} is
along the $+z$ direction. Now suppose at another instance, {\bf E}
changes to $-x$ and {\bf B} to $-y$, so the electron velocity is along
$+x$, but the Lorentz force is still along the $+z$ axis. If we have a
positive charge, the situation is the same. The fundamental reason why
we always have a positive force is because the light propagates along
the $+z$ axis and the Poynting vector is always along $+z$ and ${\bf
  v}\times {\bf B}$ points along the $+z$ axis. We test it with
various laser parameters and do not find a negative $v_z$, in
agreement with Gao \cite{gao2004}.  Under cw approximation, Rothman
and Boughn \cite{rothman2009a} gave a simple but approximate
expression for the dimensionless $v_z=\frac{1}{2}\left (
\frac{\omega_c}{\omega}\right)^2[\cos(\omega t)-1]^2$, and Hagenbuch
\cite{hagenbuch1977} gave $p_z=e^2A^2(\tau')/2mc$, both of which are
positive. Therefore, both their theories and our numerical results
agree that the axial motion of the electron is delivered by both {\bf
  E} and {\bf B}.  This is also consistent with the radiation pressure
from a laser beam can accelerate and trap particles
\cite{ashkin1970,ashkin1986}.  Figure \ref{fig2}(d) shows that $z$
reaches 1.66 $\rm \AA$ at 100 fs. Naturally, the electron can move
further as time goes by until it collides with other electrons and
ions.

\begin{figure}
  \includegraphics[angle=0,width=1\columnwidth]{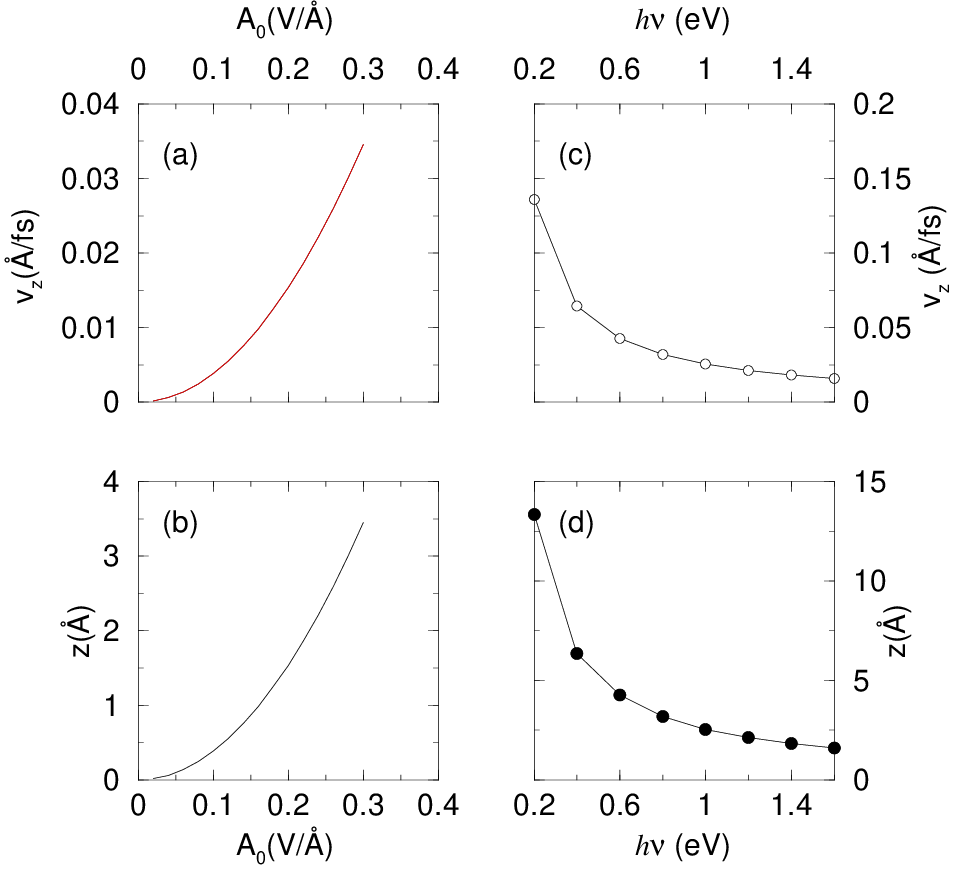}
  \caption{(a) Final velocity of the electron as a function of the
    laser amplitude $A_0$ from 0.02 to 0.30 $\rm V/\AA$, corresponding
    to the fluence $F$ from 0.09 to 20.96 \mje.  (b) Final position as
    a function of $A_0$.  (c) Final velocity of the electron as a
    function of $h\nu$.  (d) Final position of the electron as a
    function of $h\nu$. }
\label{fig3}
\end{figure}

Keeping the rest of laser parameters unchanged, i.e.,
$h\nu=1.55$ eV and $\tau=14$ fs, we increase the laser field
amplitude $A_0$ from 0.02 to 0.30 $\rm V/\AA$.  Figure \ref{fig3}(a)
shows the final velocity as a function of $A_0$.  We notice that $v_z$
change is highly nonlinear.  We fit it to a quadratic function,
$v_z=\alpha A_0^2$, where $\alpha$ is a constant of 0.383898
\an/(fsV/\an)$^2$, and find that the fit is almost perfect. Because
$|A_0|^2$ is directly proportional to the fluence, this demonstrates
$v_z$ is linearly proportional to the laser fluence, which is exactly
expected from the Poynting vector ${\bf S}={\bf E}\times {\bf
  B}/c$. Thus, both qualitatively and quantitatively our results can
be understood. Since the displacement $z$ follows the velocity, it
also increases with $A_0$ quadratically (see Fig. \ref{fig3}(b)).

With growing interest in THz pulses, we investigate the photon energy
dependence of $v_z$ and $z$.  We increase $h\nu$ from 0.2 up to 1.6
eV, while keeping both the duration $\tau=14$ fs and amplitude
$A_0=0.207\ \rm V/\AA$ fixed.  Figure \ref{fig3}(c) shows an
astonishing result: both $v_z$ and $z$ are inversely proportional to
$h\nu$.  At the lower end of $h\nu$, $v_z$ reaches 0.136 \an/fs and
$z$ reaches 13.34 $\rm \AA$ (see Fig. \ref{fig3}(d)).  Note that at
such a low amplitude, a pulse of 1.6 eV only drives the electron by
1.60 $\rm \AA$.  This explains why THz pulses become a new frontier
for ultrafast demagnetization
\cite{shalaby2015,hudl2019,lee2021}. Polley \et \cite{polley2018}
employed a THz pulse to demagnetize CoPt films with a goal toward
ultrafast ballistic switching. Shalaby \et \cite{shalaby2018} showed
that extreme THz fields with fluence above 100 mJ/cm$^2$ can induce a
significant magnetization dynamics in Co, where the magnetic field
becomes more important. \UU{ch4} {\clr Very recently, using multicycle 2.5-THz
pulses, Riepp \et \cite{riepp2024} reported incoherent and coherent
magnetization dynamics in labyrinth-type Co/Pt multilayers.} \UU{ch4end}
Our
result uncovers an important picture. When the pulse oscillates more
slowly, the electron gains more grounds. Of course, a DC current can
move electrons even further, but then it does not have enough field
intensity.  This result can be tested experimentally.

\subsection{Comparison with prior theories}

To the best of our knowledge, there has been no investigation using
the experimental laser parameters \cite{razdolski2017} as we did, so
we decide to compare our results with that of Compton scattering. This
comparison is possible because we use a free-electron model.

Assume that the initial kinetic energy of the electron is zero. The
kinetic energy gained from the photon must be equal to the loss in the
photon energy as \be \frac{1}{2}m_ev^2_e=Nh\Delta \nu,
\rightarrow v_e=\sqrt{ \frac{2Nh\Delta \nu}{m_e} } \ee where
$v_e$ is the electron velocity, $N$ is the number of photons
that interact with the electron and $\Delta \nu$ is the frequency
change computed from the wavelength change $\Delta \lambda$ as $\Delta
\nu= - \frac{c}{\lambda^2} \Delta \lambda$.  The wavelength change is
$\Delta \lambda=\frac{h}{m_ec} (1-\cos\theta) $, where $\theta$ is the
outgoing angle of the photon with respect to the incident direction
(see Fig. \ref{fig1}(b)). We directly compute the velocity of the
electron as \be v_e=\sqrt{ \frac{2Nh\Delta \nu}{m_e} }
=\frac{h}{m_e\lambda}\sqrt{2N(1-\cos\theta)}.  \ee If we use
Brorson's experimental wavelength, we find $v_e=1.1544\times 10^3
\sqrt{2N(1-\cos\theta)}$ m/s. Except that we have an
enormously large $N$, the electron velocity is around 10$^3$
m/s. Figure \ref{fig2}(c) shows the final velocity is $1.65\times
10^{-2}$ \an/fs or 1.65 $\times 10^3$ m/s. Therefore, both approaches
agree with each other very well.

\subsection{Comparison with experiments}

Although Razdolski \et \cite{razdolski2017} estimated the axial
displacement on the order of 2 nm, it is harder to compare because it
depends on the relaxation time of the electron in a particular
material.  Instead, majority of experiments measured the velocity.
Brorson \cite{brorson1987} gave $v_e=1\times 10^6$ m/s for
Au. Melnikov \cite{melnikov2011} found that a delay of 40 fs for a 50
nm of Au, so the velocity is $1.25\times 10^6$ m/s.

These velocities
are three orders of magnitude larger than our theoretical value above.
This shows that although our model does yield the correct direction of
the electron motion, the free electron model, which is sufficient for
atomic and plasma physics, is not adequate for metals, at least for
the transport property induced by laser. In the following, we discuss
several possible solutions for future research. 

\section{Beyond the free-electron model}

Given the above finding, it is very appropriate to discuss a few
alternative theories, as they are crucial to experiments
\cite{malinowski2018}.

\subsection{Drude model}

We recall first that in metals, the Drude model is often used for the
electric transport, where a voltage bias is applied along the
longitudinal direction. If an electric field $E$ (constant) is along
the $x$ direction, the electron momentum change is given as $
\frac{P_f-P_i}{\Delta t} =-eE $, which will grow with time infinitely.
Drude realized that the electron must experience collisions from other
electrons and ions, so the electron will lose its momentum, whose
value is given by $m_ev_d/\tau$. At equilibrium, these two must be
equal, so  the drift velocity is $v_d=-\frac{eE\tau}{m_e}$, where $\tau$
is the relaxation time, not to be confused with the above laser pulse
duration. $v_d$ in metals is around 0.1 m/s \cite{ohanian}.

In laser excitation, there is no such field. It is possible to include
the magnetic field into the Drude model, but as seen above, this will
lead to the same conclusion as our free electron model.

\subsection{Diffusion transport}

Diffusion transport has been proposed to understand the electron
transport. However, there is a shortcoming in these theories. They do
not specify how these initial gradients are established.  For
instance, Choi \et \cite{choi2014,choi2014b} and Fognini \et
\cite{fognini2017} replaced the particle density $n$ by the chemical
potential as $ \frac{\partial \mu_s}{\partial t}=D \frac{\partial^2
  \mu_s}{\partial z^2}-\frac{\mu_s}{\tau_s}, $ where
$\mu_s=\mu_\uparrow -\mu_\downarrow$ is the spin chemical potential
and $D$ is the spin diffusion constant, and $\tau_s$ is the spin
relaxation time.  How a nonzero $\frac{\partial^2 \mu_s}{\partial
  z^2}$ appears is not given. This is also true for the superdiffusion
model \cite{battiato2012}, where the initial velocity of the electron
is considered as an input parameter and is not possible to compare
with the experimental velocity.

\subsection{Boltzmann equation}

The third possibility is to use the Boltzmann equation. The
distribution function, $f(\br,\bv,t)$, under the influence of the laser
field,  changes as \cite{thomas1966,kittel,mahan1987} \be \frac{\partial f}{\partial
  t}=\lim_{dt\rightarrow 0} \frac{f(\br,\bv,t)-
  f(\br,\bv,t-dt)}{dt}=-\bv\cdot \nab_\br f-\dot{\bk}\cdot \nab_\bk f,
\ee where $\dot{\bk}=-\frac{e}{\hbar}(\bE+\bv \times \bB)$. We are
only interested in $k_z$ because this is along the axial
direction, $\dot{k_z}= -e v_xB_y/\hbar$. Since we already know $v_x$
and $B_y$, we can compute it easily. Figure \ref{fig4}(a) shows how $
k_z$ changes with time. We see that $\Delta k_z$ is very small, whose largest
value is around $\rm 1.5 \times 10^{-3} \AA^{-1} $. However, it is
sufficient to move electrons in solids. We take fcc Ni as an example.
Figure \ref{fig4}(b) shows the band structure of fcc Ni along the
$\Gamma$-Z direction. There are five bands across the Fermi
energy. Regardless of how small $\Delta k_z$ is, the laser field is
able to lift electrons from an occupied band to an unoccupied band, i.e.,
the intraband transition  \cite{jpcm23}. Now the questions is whether
we have enough photons. 

We can estimate the number of photons on the surface of a unit cell
from the experiments.  We take bcc Fe as an example. Its lattice
constant is $a=2.861 \rm \AA$, so its cross section is $a^2$. The
experimental fluence \cite{melnikov2011} is $F=1$ \mj and the photon
energy is $h\nu=1.55$ eV, so the number of photons per unit cell is
$3.296$. One bcc cell has two Fe atoms, so each Fe atom receives
$n=1.65$ photons. If we assume that these photons are absorbed,
$nh\nu=\frac{1}{2}m_ev^2_e$, we find $v_e=1.044 \times 10^6$ m/s. For
fcc Au, using Brorson's experimental data, we find the number of
photons per unit cell to be 5.2746. Each Au atom receives 1.32
photons. We find $v_e=0.956\times 10^6$ m/s, which agrees with the
experimental value of 10$^8$ cm/s almost perfectly.  Therefore, we
believe that it is very likely that a quantum theory, which includes
the magnetic field and the band structure, will be able to explain
experimental results quantitatively.

\section{After initial excitation: Resistive transport theory}

\begin{figure}
  \includegraphics[angle=0,width=1\columnwidth]{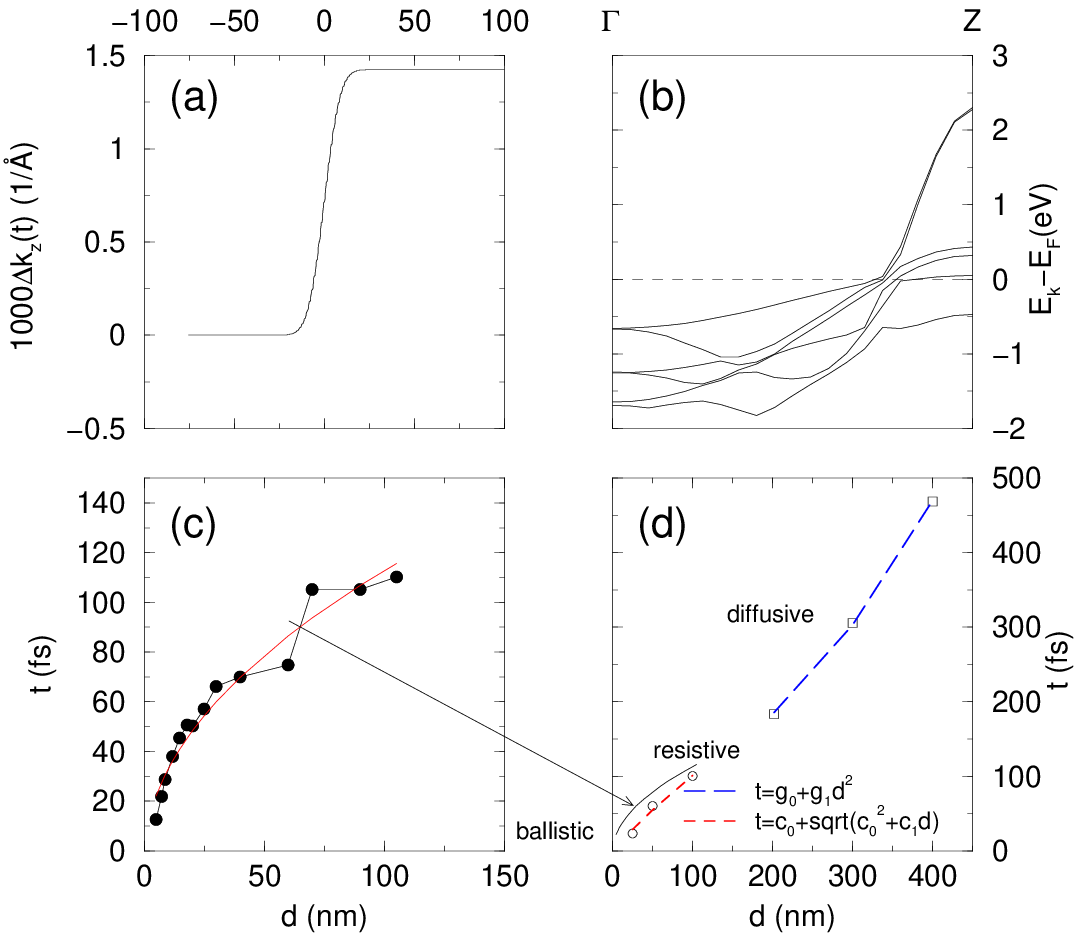}
  \caption{(a) Wavevector change $\Delta k_z$ as a function of
    time. (b) Band structure of fcc Ni along the $\Gamma$-Z axis. The
    horizontal line is the Fermi level.  (c) Photoemission data
    (filled dots) from Heckschen \et \cite{heckschen2023}. The thin
    line is our resistive theory prediction, which has a concave down.  
(d) Experimental data
    (empty circles and boxes) from Suarez \et \cite{suarez1995}. The
    short dashed line is our resistive transport fit, a typical 
concave down.  
 The long-dashed
    line is a fit to the diffusive equation. The thin solid line is
    from (c) to see how similar both experimental data are. 
}
\label{fig4}
\end{figure}

Once electrons have enough velocity, they will travel through the
materials \cite{brorson1987}. There are three different theories for
transport. Ballistic and diffusive transports are two traditional ones
\cite{knorren2002,chan2009}.  But their separation has not been very
clear.  In 1998, Bron \et \cite{bron1998} concluded that ``the
transport of quasi-electrons arriving at the peaks is neither purely
ballistic nor purely diffusive''.  This unclear region is termed the
superdiffusive transport \cite{battiato2012}, but its physics remains
unclear.  The reason is that
for a long time, there have been very few thickness
dependence experiments, besides those earlier ones
\cite{brorson1987,juhasz1993,suarez1995,melnikov2011}.  Heckschen \et
\cite{heckschen2023} carried out a systematic time-resolved two-photon
photoemission (2PPE) measurement as a function of the thickness $d$ of
Au films from 5 to 105 nm in the Au/Fe/MgO structure. The thickness of
Fe is fixed at 7 nm.  2PPE employs two laser pulses.  The pump pulse
hits on the back of the iron film through the MgO substrate and pushes
electrons toward the interface between Fe and Au layers. These hot
electrons pass through the interface and enter the Au layer. Once they
reach the surface of Au, a probe pulse of 4 eV knocks them
out. Experimentally, Heckschen \et collected the electrons as a
function of kinetic energy after each time delay between the pump and
probe.  At $E-E_F=0.6$ eV, where $E$ is the electron energy and $E_F$
is the Fermi energy, they obtained an accurate dependence of the
reduced time delay $t^*$ (see their paper for detail) as a function of
$d$.  We use Webdigitizer \cite{web} to read off their experimental
data from their Fig. 4 and reproduce them in Fig. \ref{fig4}(c),
without error bars.  According to their analysis, they found that the
dependence of time $t$ on $d$ is sublinear, and concluded that the
velocity follows a linear behavior.

However, this concave curve looks strangely familiar to us. In
particular, it is not linear, in contrast to the wavelike heat
transport \cite{sobolev2022}. Instead it is more like a function as
\be t=c_0 +\sqrt{c_0^2+c_1d}, \label{t1} \ee where we use $t$ instead
of their $t^*$ for simplicity, $c_0$ and $c_1$ are two fitting
parameters. A quick test indeed confirms our guess, where
$c_0=-4.14678$ fs and $c_1=136.398$ fs$^2$/nm. The fitted data is the
solid red line in Fig. \ref{fig4}(c).  The match is excellent, given
that we only have two fitting parameters and have the experimental
errors.  The reader must wonder why we only need two fitting
parameters, instead of three, which would appear more natural. This is
because we do not have a prior bias toward ballistic, superdiffusive
or diffusive transport. We just want to see where the experiment leads
us to.  We can rewrite Eq. \ref{t1} as \be d=-\frac{2c_0}{c_1}t
+\frac{1}{c_1}t^2,\label{t2} \ee which is a parabola.

This reminds us the motion with a constant acceleration. Consider a
one-dimensional motion. The displacement is \be \Delta x =v_i t
+\frac{1}{2} at^2,\ee where $a$ is the acceleration, $\Delta x$ is the
same as $d$ above, $v_i$ is the initial velocity.  $t$ is the time
which may differ from the experimental one (see above and also below).
However, in the photoemission, electrons are collected at a fixed
kinetic energy, so its final velocity $v_f$ remains the same. We need
to replace $v_i$ by $v_f$ as \be \Delta x=(v_f-a t)t+\frac{1}{2}
at^2=v_ft -\frac{1}{2} at^2,\label{t3} \ee which is a crucial step as
will be seen below.  Comparing Eqs. \ref{t2} and \ref{t3}, we find
$a=-2\frac{1}{c_1}=-0.00733 \rm nm/fs^2 <0$.  This  suggests
that electrons inside the Au film deaccelerate under a uniform drag
from the ions and other electrons. So it is neither ballistic, nor
superdiffusive, nor diffusive. We call it electron resistive transport. 
The final velocity,
$v_f=-\frac{2c_0}{c_1}=0.0608$ nm/fs, appears too small if we take 0.6
eV as the electron kinetic energy, probably because their time $t^*$
is not an absolute time, as they called it the propagation time
\cite{heckschen2023}. A single agreement is not enough to establish
that the uniform drag is indeed the underlying mechanism for electron
transports under laser excitation. We move on to the Brorson's data
\cite{brorson1987}, but unfortunately there is not enough data within
150 nm (they only have two points).

Fortunately, Suarez \et
\cite{suarez1995} have three data points below and above 150 nm. We
obtain the data using Webdigitizer \cite{web} from the inset of their
Fig. 1 and reproduce their data in \ref{fig4}(d) (see three empty
circles and boxes).  In their original paper, they linked all six data
points together and used a straight line to fit their data to suggest
a possible ballistic transport, because they noticed that the time
``does not vary as the second power of the film thickness as one would
expect from random-walk thermal diffusion''. The red short-dashed line
is our fitted curve $t=c_0+\sqrt{c_0^2+c_1d}$, where $c_0=-239.382$ fs
and $c_1=585.481$ fs$^2$/nm.  The fitting is quite good as well, given
that they only have three data points.  What is more important that it
has the same trend as Heckschen's data (the thin solid line from
Fig. \ref{fig4}(c) is reproduced in Fig. \ref{fig4}(d)).  Caution must
be taken to make a quantitative comparison. Suarez's time is the
traversal time which is defined as the transient reflectivity reaches
15\%.  From $c_0$ and $c_1$, we find $a=-0.0034 $ nm/fs$^2$ and
$v_f=0.82$ nm/fs. We see both accelerations have the same magnitude,
but their velocities differ. Suarez's velocity is closer to the Fermi
velocity. More experimental data are necessary.

To investigate whether our theory can reproduce the ballistic limit,
we rewrite Eq. \ref{t3} as \be \frac{1}{2}at^2-v_ft+\Delta x=0, \ee so
the time $t$ is given by \be t(\Delta x)=\frac{v_f\pm
  \sqrt{v_f^2-2a\Delta x}}{a}, \ee which reveals immediately why we
must only have two fitting parameters for a $\Delta x$.  Since $v_f$
and $\Delta x$ are both positive, we only choose a root for a positive
$t$, which avoids the complex math if we used $v_i$.  In our resistive
transport, $a<0$, so the numerator must be negative as well. Therefore
we choose a negative sign in the numerator, \be t(\Delta
x)=\frac{v_f- \sqrt{v_f^2-2a\Delta x}}{a}. \label{t5} \ee To reproduce
the ballistic limit, we expand the square root in Eq. \ref{t5} by
using $\sqrt{1\pm x}=1\pm \frac{1}{2}x
-\frac{1}{2}\frac{1}{4}x^2+\cdots$ to find, \be t(\Delta
x)=\frac{\Delta x}{v_f} + \frac{a(\Delta x)^2}{2v_f^3}+\cdots.  \ee If
we drop the second and other higher order terms, we get the ballistic
limit $ t(\Delta x)=\frac{\Delta x}{v_f}$.  According to the anomalous
multiphoton photoelectric measurement \cite{kupersztych2005}, the
photoelectron current has a dramatic enhancement when the gold
overlayer thickness is nearly 43 nm, i.e., right in the middle of the
resistive transport. This anomalous effect is likely due to the
resistive transport (see Fig. \ref{fig4}(d)). Interestingly,
Kupersztych and Raynaud \cite{kupersztych2005} already stated that
``electromagnetic energy is transferred ... to conduction electrons in
a duration less than the electron energy damping time''. This damping
now shows up in our negative acceleration.

We can estimate how much the electron velocity is reduced. Using
$a=-0.0034 $ nm/fs$^2$, every additional 100 fs, the velocity is
reduced by 0.34 nm/fs. As the electrons slow down, they start to
accumulate spatially, ready for a normal diffusion.  The resistive
transport transitions to the regular diffusion (see
Fig. \ref{fig4}(d)).  We fit Suarez's data (last three points) to
$t(d)=g_0+g_1d^2$, where $g_0=88.2088$ fs and $g_1=0.00238099$
fs/nm$^2$. A spatial separation between the resistive and diffusive
transports in Au is around 200 nm as seen from Fig. \ref{fig4}(d).
The reason why our simple resistive picture works well is because the
interband state in Au is roughly 3 eV above $E_F$
\cite{suarez1995,liu2005} and the transport is strictly due to
intraband transition \cite{jpcm23}. This may be different for
different materials \cite{persson2016}. We are also aware of a similar
concave down in the size dependence of the electron thermalization
time for nanoparticles \cite{voisin2000,voisin2001}, but the
dependence is unlikely related to the electron velocity, since its
value is too small.

\section{Conclusion}

This study centers on two key themes. First we have shown that the
electron axial transport is the joint effect of the electric and
magnetic fields of the laser pulse. Each field alone cannot lead to
the transport along the axial direction. The electric field provides a
strong transverse velocity, while the magnetic field steers the
electron moving along the light propagation direction.  This welcoming
result requires a nonzero {\bf B}, but a nonzero {\bf B} subsequently
requires a spatially dependent vector potential ${\bf A}(\br,t)$,
owing to ${\bf B}=\nabla \times {\bf A}(\br,t)$.  The dipole
approximation is frequently employed in model and first-principles
theories \cite{jap11,krieger2015,chen2019w,prl20}, but in order to
describe electron transports in thin films, one has to adopt ${\bf
  A}(\br,t)$, not ${\bf A}(t)$.  However, quantitatively, our
free-electron model is still unable to reproduce the experimental
velocity.  Going beyond the free electron model, if all the photon
energies are absorbed, we can quantitatively reproduce the
experimental velocity. Second, by carefully analyzing the experimental
data, we propose a concept of the resistive transport, which unifies
the ballistic and superdiffusive transports. Here electrons, after
initial acceleration by the laser pulse, deaccelerate under a uniform
resistance from other electrons and ions.  Earlier part of the
resistive transport is the ballistic transport, while the later part
covers superdiffusive transport. As electrons slow down, they enter a
normal diffusion regime.  Our study provides a much simpler picture on
the ultrafast time scale, and reveals a big deficiency with the
existing theories \cite{mplb18,aip20,liu2020,liu2023,kang2023}.
\UU{ch3}{\clr A new experiment is now available in copper. Jechumtal
  \et \cite{jechumtal2024} unknowingly reported a similar square-root
  dependence. They thought that because of the limitation of their
  model, the model does not capture the slightly nonlinear trend in
  their Figs. 3(a) and 3(b) for the thickness $d<3$ nm. They even put
  a shaded triangle over the nonlinear portion of the curve. Now, this
  is in fact the result of the resistive transport, another proof of
  our theory.}

\acknowledgments

We appreciate the helpful communications with Dr. Uwe Bovensiepen
(University of Duisburg-Essen, Germany) and Dr. David G. Cahill
(University of Illinois).  This work was partially supported by the
U.S. Department of Energy under Contract No. DE-FG02-06ER46304. Part
of the work was done on Indiana State University's high performance
Quantum and Obsidian clusters.  The research used resources of the
National Energy Research Scientific Computing Center, which is
supported by the Office of Science of the U.S. Department of Energy
under Contract No. DE-AC02-05CH11231.

$^*$ guo-ping.zhang@outlook.com.
 https://orcid.org/0000-0002-1792-2701

The data that support the findings of this study are available from
the corresponding author upon reasonable request.

\clearpage

\end{document}